\documentclass[amsfonts, amssymb, amsmath, reprint, showkeys, nofootinbib, twoside, prx]{revtex4-1}

\usepackage{amsthm}
\usepackage{mathtools}
\usepackage[italicdiff]{physics}
\usepackage{xcolor}
\usepackage{graphicx}
\usepackage[left=23mm,right=13mm,top=35mm,columnsep=15pt]{geometry} 
\usepackage{adjustbox}
\usepackage{placeins}
\usepackage[T1]{fontenc}
\usepackage{lipsum}
\usepackage{csquotes}

\newcommand*\chem[1]{\ensuremath{\mathrm{#1}}}
\usepackage{tikz}
\usetikzlibrary{matrix,fit,positioning}
\usetikzlibrary{decorations.pathmorphing}
\tikzset{
mymat/.style={
    matrix of math nodes,
    left delimiter=|,right delimiter=|,
    align=center,
    column 1/.style={align=center},
    row sep = -\pgflinewidth,
  },
mymats/.style={
    mymat,
    nodes={draw}
}  
}

\newcommand{\epsa}{\varepsilon_0}
\newcommand{\epsb}{\varepsilon_1}

\begin{document}

\newcommand{\trel}{t_\mathrm{rel}}
\newcommand{\tave}{t_\mathrm{ave}}
\newcommand{\kk}{\mathbf{k}}
\renewcommand{\qq}{\mathbf{q}}
\newcommand{\pp}{\mathbf{p}}
\newcommand{\ea}{\epsa}
\newcommand{\eb}{\epsb}

\title{Observing coherences with time-resolved photoemission}

\author{Alexander F. Kemper}
 \email{akemper@ncsu.edu}
\affiliation{Department of Physics,
 North Carolina State University,
 Raleigh, NC
}

\author{Avinash Rustagi}
\affiliation{School of Electrical and Computer Engineering, Purdue University, West Lafayette, IN 47907}

\date{\today}

\begin{abstract}
We discuss the potential creation and measurement of coherences in both dispersive solids and qubit-like single levels using
current generation time- and angle-resolved photoemission technology.  We show that in both cases, when both the
pump and the probe overlap energetically with the coherent levels, and when the probe preferentially measures one
level as compared to the other, that the time-resolved photoemission signal shows a beating pattern at the energy difference
between the levels. In the case of dispersive bands, this leads to momentum-dependent oscillations, which may be used
to map out small energy scales in the band structure.  We further develop the two-sided Feynman diagrams for time-resolved
photoemission, and discuss the measurement of decoherence to gain insight into the characteristics of qubit and dispersive
bands.
\end{abstract}

\maketitle

\vspace{1em}
\noindent{\bf {\large Main}}\\

Coherences are fundamental to quantum mechanics.
One particularly
striking example is quantum computing, where coherences between the two qubit states
are foundational to the concept. Decoherence, which occurs naturally
as qubits interact with their environment, is a strong limiting factor on the use of quantum
computation. Thus, there is a drive to understand the nature of the decoherence in order to 
improve the quality of existing qubit technology, and for the development of new qubit candidates.\cite{DOEQUANTUMPLATFORM}

A second area where coherence plays an important role is
condensed matter physics, in emergent phenomena
that arise out of
the interaction
between constituents.
This field is rife with small energy scales: from small
energy gaps induced by emergent phenomena\cite{hashimoto2014energy} or Kondo physics\cite{demsar2006photoexcited} to magnetic effects, and we are often limited by our ability to resolve physics at the smallest energy scales. This has
led to an ever increasing development of higher resolution experiments to explore the fundamental physics at play.

Among the developments in the quest to understand coherence in solids,
time-resolved spectroscopy, and in particular time- and angle-resolved photoemission
spectroscopy (tr-ARPES) was envisioned to be able to resolve
very small energy differences by working in the time domain. 
In certain cases, this was indeed possible; oscillations due to coherent phonons were observed \cite{gerber2017femtosecond,hein2019mode,yang2019mode,zhang2020coherent} and used to infer
properties of interactions.
However, the general theme of utilizing the inverse relationship between time and energy
to resolve small energy scales has been limited because, while the scales
do obey an inverse relationship, the response of driven systems is markedly different
from their equilibrium response\cite{kemper2018general,RevModPhys.86.779}. Furthermore,
the dynamics observed are often limited to populations (see e.g. Refs.~\cite{rameau_energy_2016,konstantinova2018nonequilibrium,freutel2019optical}) which do not
typically reflect coherences. 

On the other hand, in the study of molecules (or solids with strong resonances), the time domain
optical response show both populations and coherences.
Nonlinear optical spectroscopy (specifically multi-dimensional spectroscopy),
maps the beating patterns in the temporal response onto populations and coherences\cite{boyd2003nonlinear}.
The difficulty in applying multi-dimensional spectroscopies 
to solids is largely because, in contrast to molecules, solids do not typically 
have a finite set of strong resonances (although exciton complexes and semiconductor
nanostructures are notable exceptions\cite{moody2017advances,kandada2019perspective,thouin2019enhanced}),
and optical measurements average over the Brillouin zone. This makes the spectra difficult
to interpret without a large amount of prior knowledge about the solid (i.e. its energy levels, dipole matrix elements, and oscillator strengths)\cite{smallwood2018multidimensional}. This limits the usefulness of this approach for studying less well-known, potentially complex interacting systems.

Here, we bring the concept of coherences from non-linear optical
spectroscopy to photoelectron spectroscopy. This technique, based on current generation tr-ARPES experiments can bring new insights into the two fields.  For the strong resonances involved
in qubits, tr-ARPES can enable precise characterization
of the qubits and the coherences. In solids, coherences between bands can be produced and measured,
yielding beat frequencies corresponding to the energy difference. First, this has an advantage over
optical measurements due to its momentum selectivity, and secondly it truly embodies the inverse relationship
between time and energy resolutions. Along with the concepts, we will introduce the two-sided
Feynman diagrams that are commonly
used in non-linear optics\cite{yee1977diagrammatic,yee1978diagrammatic,boyd2003nonlinear} as a tool to keep track of the pathways before and after the photoemission process. This language naturally captures the coherence
and the chronological progress as the system undergoes photoexcitation by the pump, time evolution,
and photoelectron emission.

\vspace{1em}
\noindent{\bf {\large Results.}}\\

\begin{figure}[h]
\resizebox{0.4\columnwidth}{!}{
\begin{tikzpicture}
	\draw (-2,-1) node[anchor=east] {$\lbrace \ket{c}\rbrace$} -- (1,-1) ;
	\draw [double] (-2,0) node [anchor=east] {$\ket{0}/\ket{1}$} -- (1,0);
	\draw (-2,1) node [anchor=east] {$\ket{v}$} -- (1,1);
	
	\draw [red, very thick, ->] (-1,-1) .. controls  (-1.2,-0.5) .. (-1,0);
	\draw (-1,-.5) node [anchor=west] {Pump};

	\draw [blue, very thick, ->] (0,0) .. controls  (-0.2,0.5) .. (-0,1);
	\draw (0,.5) node [anchor=west] {Probe};
\end{tikzpicture}
}
\resizebox{0.49\columnwidth}{!}{
\begin{tikzpicture}[font=\sffamily,
    every left delimiter/.style={xshift=.7em},
    every right delimiter/.style={xshift=-1em}]
\matrix[mymat] at (3,0) (mat1)
{   $\dyad{v}{v}$ \\
    $\dyad{v}{1}$ \\
    $\dyad{0}{1}$ \\
    $\dyad{c}{c}$ \\
};
\draw[stealth-,color=red] (mat1-3-1.south -| mat1.east) -- ++(4mm,-4mm);

\draw[-stealth,decorate,    color=blue, decoration={
    zigzag,
    segment length=4,
    amplitude=.9,post=lineto,
    post length=2pt
}] (mat1-1-1.south -| mat1.east) -- ++(4mm,4mm) 
node[color=black,text width=2cm,align=left,pos=0.9,above right]{$\cos\theta \hat{c_0}(t')+ \sin\theta\hat{c_1}(t')$};

\draw[stealth-,color=red] (mat1-3-1.south -| mat1.west) -- ++(-4mm,-4mm) node[pos=1.2,below,color=black]{$t=t'=0$};

\draw[-stealth,decorate,color=blue, decoration={
    zigzag,
    segment length=4,
    amplitude=.9,post=lineto,
    post length=2pt
}] (mat1-2-1.south -| mat1.west) -- ++(-4mm,4mm) 
node[color=black,text width=2cm,align=left,pos=0.2,above left]
{$\cos\theta \hat{c_0}(t) + \sin\theta\hat{c_1}(t)$};

\end{tikzpicture}
}
\caption{{\bf Energy levels and two-sided Feynman diagrams for a qubit in a solid.} 
Left: Energy levels and transitions for the 2-state toy model. The pump produces a coherence (e.g. $\dyad{0}{1}$), which
is photoemitted by the probe.  Right: corresponding two-sided Feynman diagram showing the pump/probe
process through the intermediate coherences (see text for description).}
\label{fig:3state}
\end{figure}
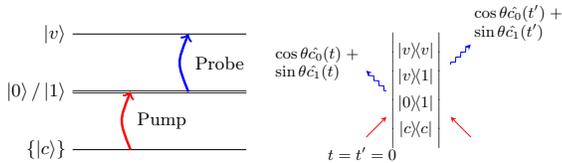

\noindent{\bf Photocurrent from a coherence.}
To begin, we consider a purely off-diagonal density matrix, i.e. one composed of {\it coherences}.
To simplify notation, and to make a direct connection to qubits,
we study
a 2-state system as shown in Fig.~\ref{fig:3state}. 
The 2-state system has two initially unoccupied levels $\ket{0}$ and $\ket{1}$, close by in energy. We
augment this model with two ancillary states: 
some deeper lying occupied levels $\lbrace{\ket{c}\rbrace}$ that act as a source electrons,
and the empty state $\ket{v}$ (vacuum) where
the electron from $\ket{0}/\ket{1}$ ejects into a free electron state outside the crystal. 
A coherence may be produced by a pump whose 
linewidth overlaps the energies
of $\ket{0}$ and $\ket{1}$ ($\epsa$ and $\epsb$, respectively),
\begin{align}
\rho_\mathrm{coh} \propto \mu_{c0} \mu_{1c} \ketbra{0}{1}  +\mu_{0c} \mu_{c1} \ketbra{1}{0}
\end{align}
where $\mu_{0c}$ is the transition matrix element from $\ket{c}$ to $\ket{0}$, and similar for $\mu_{1c}$.

For photoelectrons ejected from $\ket{0}$, we may evaluate the photocurrent through Eq.~\ref{eq:gless_from_rho}
using the operator 
$\hat c_0$ which produces a transition from $\ket{0}$ to $\ket{v}$.  It is clear that
this yields $0$ for the photocurrent; the operators produce transitions
between $\ket{0}$ and $\ket{v}$, but the final trace yields nil since $\ket{0}$ and $\ket{1}$ are orthogonal
and $\rho_\mathrm{coh}$ only contains the cross-terms; 
the same result is obtained for the photocurrent from $\ket{1}$.

There is, however, an intriguing case where this expectation value is nonzero, namely
if the photoemission process occurs with finite likelihood for both
states $\ket{0}$ and $\ket{1}$. One common reason this may occur is if
$\ket{0}$ and $\ket{1}$ are composites of two different states,
e.g. $\ket{\alpha}$ and $\ket{\beta}$, and the photoemission is more sensitive to one or the other due to
matrix elements. This latter case may be common in solids where the bands are composed of multiple
orbitals of different kinds, e.g. in $p$-$d$ or $p$-$f$ systems such as transition metal oxides or $f$-electron
materials such as \chem{SmB_6}. In these materials, the orbitals are different in spatial extent, and 
disparate matrix elements may occur. Qubit candidates, in particular those embedded in solids,
can have similar properties.\cite{gordon2013quantum,hays2019continuous,gottscholl2020initialization,borjans2020resonant,lane2020integrating}

\begin{figure}[htpb]
	\includegraphics[width=0.49\textwidth, clip=true, trim=0 0 0 30]{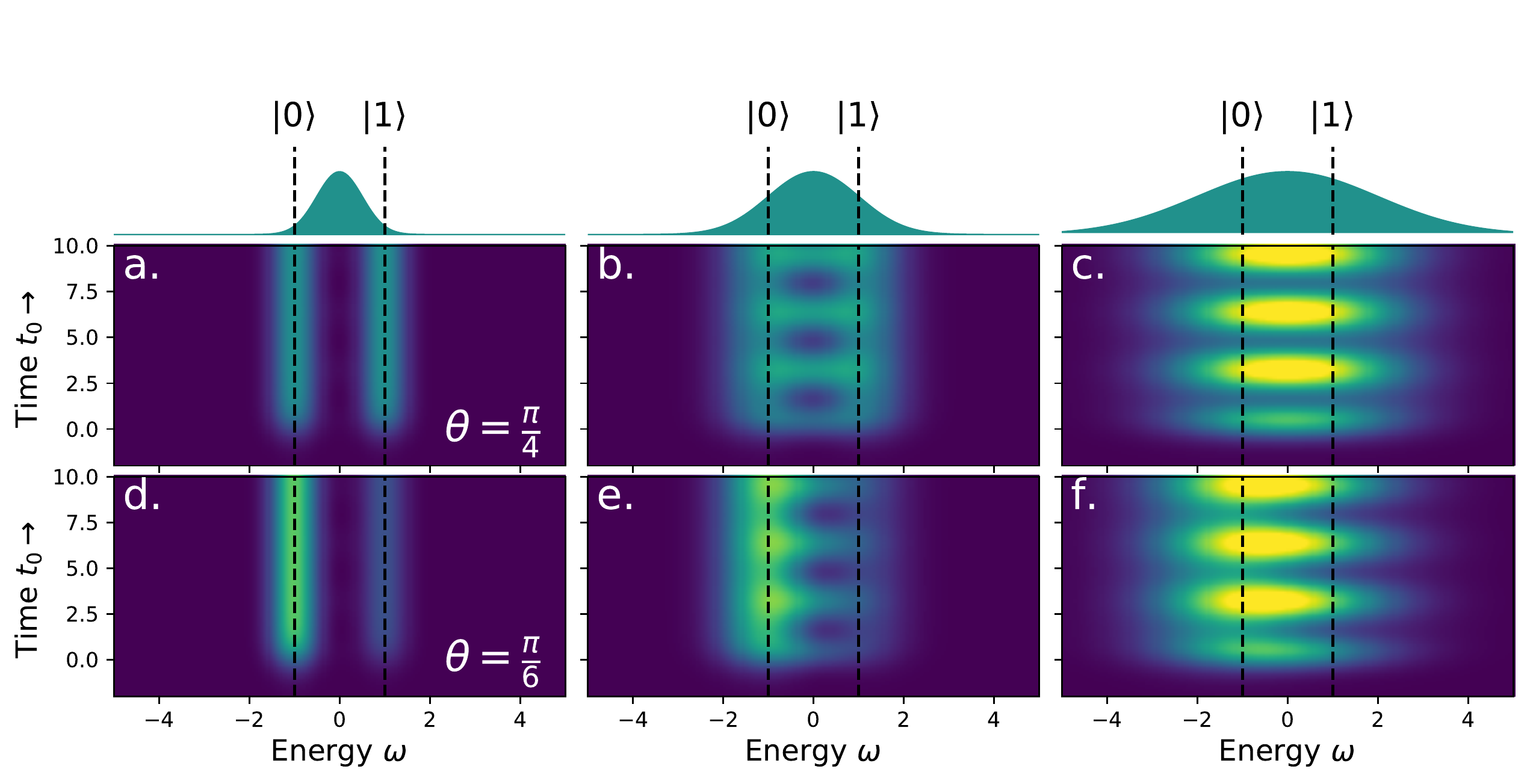}
	\caption{{\bf Time-resolved photoemission from a population with a coherence.}
	Panels {\bf a.}-{\bf c.} show photoemission 
	with equal probabilities of emission ($\theta=\pi/4$) and, and panels {\bf d.}-{\bf f.} show preferential photoemission of the lower
	energy state ($\theta=\pi/6$). The three columns represent three values of the probe linewidth
	$\sigma_\omega$ as illustrated diagrammatically above.}
	\label{fig:simple_arpes}
\end{figure}

When one of these scenarios occurs, may we replace the $\hat c_0$ photoemission
operator with a combination $\cos(\theta) \hat c_0 + \sin(\theta) \hat c_0$ (and similar for the creation operators)
that conveys the mixed character of the states, or in another way, the
photoemission matrix elements. 
First, let us consider the new photoemission operators
as they act on an incoherent, population-only density matrix
\begin{align}
\rho_\mathrm{pop} = |\mu_{c0}|^2 \dyad{0}{0} + |\mu_{c1}|^2 \dyad{1}{1}.
\end{align}
which yields the lesser Green's function
\begin{align}
G^{<\mathrm{,pop}}(t,t') = &i  |\mu_{c0}|^2 \cos(\theta)^2 e^{-i \epsa\left(t-t'\right)} \\
+ &i |\mu_{c1}|^2 \sin(\theta)^2 e^{-i \epsb\left(t-t'\right)}.
\end{align}
As expected, we find a contribution from each state separately, with their
appropriate photoemission matrix elements.
Moving to the coherent density matrix, evaluating Eq.~\ref{eq:gless_from_rho}
with the restriction that the matrix elements are entirely real or imaginary yields
\begin{align}
G^{<,\mathrm{coh}}(t,t')  =& i \sin(2\theta) 
\mu_{c0} \mu_{1c} e^{-i\left(\frac{\epsa + \epsb}{2}\right)t_\mathrm{rel} } \nonumber\\
&\times \cos(\left(\epsa - \epsb\right)t_\mathrm{ave}) ,
\label{eq:Glesscoh_post}
 \end{align}
where we have rotated to relative time $\trel$ and average (measurement) time $\tave$.
Recalling that the expression for the photocurrent involves a windowed Fourier transform over $t_\mathrm{rel}$,
we conclude that we can find a peak in the photocurrent midway between the
energies $\epsa$ and $\epsb$, which oscillates in average time with the corresponding
beat frequency $\omega_\mathrm{beat}=\epsa -\epsb$. In this simplest form,
this illustrates the existence of a beating pattern in time-resolved
photoemission measurements due to coherence between two electronic states, which may be seen
when we consider the full signal in Fig.~\ref{fig:simple_arpes} (below).

\vspace{1em}
\noindent{\bf Two-sided Feynman diagrams for photoemission.}
To economically evaluate the contribution of populations and coherences,
we introduce an extension of the two-sided Feynman diagrams used in non-linear spectroscopy\cite{yee1977diagrammatic,yee1978diagrammatic,boyd2003nonlinear}
to time-resolved ARPES. 
These are developed by considering the
creation and annihilation operators as applying to one side of the density matrix or the other,
and conveying the process diagrammatically. 
Due to the cyclic invariance of the trace Eq.~\ref{eq:gless_from_rho} 
may be rewritten as
\begin{align}
G_\kk^<(t,t') 
=  i \Tr & \left\lbrace  \hat U(t_f,t) \hat c_\kk \hat U(t,0) \rho \right. \nonumber \\ 
&\times \left. \hat U(0,t') \hat c_\kk^\dagger \hat U(t',t_f)
\right \rbrace,
\label{eq:gless_for_diagrams}
\end{align}
where we have introduced an arbitrary final time $t_f$.
Viewing the expectation value
as operators and time evolution
acting on the two sides of a density
matrix suggests that the operators and time evolution
may be represented in a similar manner as those used in nonlinear optics\cite{yee1977diagrammatic,yee1978diagrammatic,boyd2003nonlinear}. 
When applied to the 2-level system under discussion, the resulting diagram for the coherence $\dyad{0}{1}$ is
shown in Fig.~\ref{fig:3state}. The diagram is read from bottom to top, where the system evolves under the Hamiltonian
during each interval; accordingly, if picks up a phase factor $\exp\left(-i (\varepsilon_a -\varepsilon_b)(\Delta t)\right)$ if it
resides in the state $\dyad{0}{1}$ for an interval $\Delta t$.
The pump creates the coherence at time
$t=t'=0$; for simplicity, here we will consider the
process that occurs on sufficiently short times, i.e.
within the pump pulse, and that the coherence does not
time evolve during the pump.
Following that, the annihilation operator acts on the left $\ket{}$ side at time $t$,
and produces a transition from $\ket{0}$ to $\ket{v}$. Since
the annihilation operator is a mixture of $\ket{0}$ and $\ket{1}$, 
the process picks up a factor $\cos\theta$. A similar process occurs
on the right $\bra{}$ side with the creation operator at time $t'$. Altogether we collect terms to find
\begin{align}
G^{<\mathrm{,coh}}(t,t') = -i & \cos(\theta)\sin(\theta)\mu_{c0}\mu_{1c} \times\nonumber\\
\left[  e^{-i(\epsa - \epsb)t}  \right. & \left.
 e^{i\epsb(t'-t)} +
 e^{-i(\epsb - \epsa)t}e^{i\epsa(t'-t)}
\right].
\end{align}
Simple manipulations show that this is identical to 
Eq.~\ref{eq:Glesscoh_post}, and that these diagrams are thus a faithful representation
of the process.

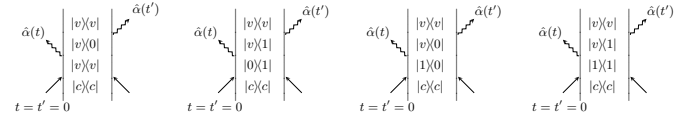
\begin{figure}[h]
\resizebox{0.5\textwidth}{!}{
\begin{tikzpicture}[font=\sffamily,
    every left delimiter/.style={xshift=.4em},
    every right delimiter/.style={xshift=-.4em}]
\matrix[mymat] at (0,0) (mat1)
{   $\dyad{v}{v}$ \\
    $\dyad{v}{0}$ \\
    $\dyad{v}{v}$ \\
    $\dyad{c}{c}$ \\
};
\draw[stealth-] (mat1-3-1.south -| mat1.east) -- ++(4mm,-4mm);
\draw[-stealth,decorate, decoration={
    zigzag,
    segment length=4,
    amplitude=.9,post=lineto,
    post length=2pt
}] (mat1-1-1.south -| mat1.east) -- ++(4mm,4mm) node[pos=0.9,above right]{$\hat \alpha(t')$};
\draw[stealth-] (mat1-3-1.south -| mat1.west) -- ++(-4mm,-4mm) node[pos=1.2,below]{$t=t'=0$};
\draw[-stealth,decorate, decoration={
    zigzag,
    segment length=4,
    amplitude=.9,post=lineto,
    post length=2pt
}] (mat1-2-1.south -| mat1.west) -- ++(-4mm,4mm) node[pos=0.75,above left]
{$\hat \alpha(t)$};

\matrix[mymat,right=3cm of mat1] (mat2)
{   $\dyad{v}{v}$ \\
    $\dyad{v}{1}$ \\
    $\dyad{0}{1}$ \\
    $\dyad{c}{c}$ \\
};
\draw[stealth-] (mat2-3-1.south -| mat2.east) -- ++(4mm,-4mm) ;
\draw[-stealth,decorate, decoration={
    zigzag,
    segment length=4,
    amplitude=.9,post=lineto,
    post length=2pt
}] (mat2-1-1.south -| mat2.east) -- ++(4mm,4mm) node[pos=0.75,above right]
{$\hat \alpha(t')$};
\draw[stealth-] (mat2-3-1.south -| mat2.west) -- ++(-4mm,-4mm) node[pos=1.2,below]{$t=t'=0$};
\draw[-stealth,decorate, decoration={
    zigzag,
    segment length=4,
    amplitude=.9,post=lineto,
    post length=2pt
}] (mat2-2-1.south -| mat2.west) -- ++(-4mm,4mm) node[pos=0.75,above left]
{$\hat \alpha(t)$};

\matrix[mymat,right=3cm of mat2] (mat3)
{   $\dyad{v}{v}$ \\
    $\dyad{v}{0}$ \\
    $\dyad{1}{0}$ \\
    $\dyad{c}{c}$ \\
};
\draw[stealth-] (mat3-3-1.south -| mat3.east) -- ++(4mm,-4mm) ;
\draw[-stealth,decorate, decoration={
    zigzag,
    segment length=4,
    amplitude=.9,post=lineto,
    post length=2pt
}] (mat3-1-1.south -| mat3.east) -- ++(4mm,4mm) node[pos=0.75,above right]
{$\hat \alpha(t')$};
\draw[stealth-] (mat3-3-1.south -| mat3.west) -- ++(-4mm,-4mm) node[pos=1.2,below]{$t=t'=0$};
\draw[-stealth,decorate, decoration={
    zigzag,
    segment length=4,
    amplitude=.9,post=lineto,
    post length=2pt
}] (mat3-2-1.south -| mat3.west) -- ++(-4mm,4mm) node[pos=0.75,above left]
{$\hat \alpha(t)$};

\matrix[mymat,right=3cm of mat3] (mat4)
{   $\dyad{v}{v}$ \\
    $\dyad{v}{1}$ \\
    $\dyad{1}{1}$ \\
    $\dyad{c}{c}$ \\
};
\draw[stealth-] (mat4-3-1.south -| mat4.east) -- ++(4mm,-4mm) ;
\draw[-stealth,decorate, decoration={
    zigzag,
    segment length=4,
    amplitude=.9,post=lineto,
    post length=2pt
}] (mat4-1-1.south -| mat4.east) -- ++(4mm,4mm) node[pos=0.75,above right]
{$\hat \alpha(t')$};
\draw[stealth-] (mat4-3-1.south -| mat4.west) -- ++(-4mm,-4mm) node[pos=1.2,below]{$t=t'=0$};
\draw[-stealth,decorate, decoration={
    zigzag,
    segment length=4,
    amplitude=.9,post=lineto,
    post length=2pt
}] (mat4-2-1.south -| mat4.west) -- ++(-4mm,4mm) node[pos=0.75,above left]
{$\hat \alpha(t)$};

\end{tikzpicture}
}
\caption{Full set of diagrams for the photoemission from the set of states $\{\ket{0},\ket{1}\}$. The $\hat \alpha$
operators are defined as $\hat \alpha = \cos\theta\ \hat c_0 + \sin\theta\ \hat c_1$.}
\label{fig:fullset}
\end{figure}
\vspace{1em}
\noindent{\bf {Coherences in qubits}}
\noindent Using the two-sided Feynman diagrams, it is straightforward to evaluate the full set
of contribution including both pieces. The full set of diagrams is shown in Fig.~\ref{fig:fullset}. 
\begin{align}
G^<(t,t') = &
i\rho_0^2 e^{-i\epsa\left(t-t'\right)} +i\rho_1^2 e^{-i\epsb\left(t-t'\right)} \nonumber\\
+&i\rho_0 \rho_1 \left(e^{-i\epsa t + i\epsb t'} + e^{-i\epsb t + i\epsa t'}  \right),
\end{align}
where $\rho_0=\mu_{ca} \cos(\theta)$ and $\rho_1 = \mu_{cb} \sin(\theta)$.
We can use this expression to decompose the Fourier transform used to evaluate the photocurrent
Eq.~\ref{eq:fourier} using a Gaussian probe profile with width $\sigma_t$,
\begin{align}
    \mathcal{I}(\omega,t_0) &=\left\lvert 
    \rho_0 e^{-i \epsa t_0} g_0(\omega) + \rho_1 e^{-i \epsb t_0} g_1(\omega )
    \right\rvert^2 \nonumber \\
    &= \rho_0^2 g_0(\omega)^2 + \rho_1^2 g_1(\omega)^2 \nonumber \\
    &+ \rho_0 \rho_1 g_0(\omega) g_1(\omega) \cos\left[ \left(\epsa-\epsb\right) t_0\right],
    \label{eq:exact}
\end{align}
where $g_{0/1}(\omega)$ are a Gaussian functions of width $\sigma_t^{-1}\equiv
\sigma_\omega$ centered around $\varepsilon_{0/1}$. This expression shows that the spectrum is positive
definite, and has the expected incoherent contributions at the energies of the individual levels.
We may rewrite the product $g_0(\omega) g_1(\omega)$ in the interference term as
\begin{align}
g_0(\omega)g_1(\omega) 
&= e^{-\frac{1}{\sigma_\omega^2}\left(\omega - \frac{\epsa+\epsb}{2}\right)^2}e^{ - \frac{1}{4\sigma_\omega^2}\left(\epsa-\epsb\right)^2},
\end{align}
and see that this term suppress the interference once the separation between $\epsa$ and $\epsb$ becomes
large on the scale of the energy resolution $\sigma_\omega$. It also indicates that the interference signal appears halfway between 
the two levels, in agreement with the earlier simpler analysis (Eq.~\ref{eq:Glesscoh_post}).

The resulting photoemission intensities from Eq.~\ref{eq:exact} for several values of probe width $\sigma_\omega$
and mixing angle $\theta$ are
shown in Fig.~\ref{fig:simple_arpes}. To simulate the pumping process we have
applied a smooth cutoff at $t=0$. For equal probabilities of photoemission ($\theta=\pi/4$) from both
states (Fig.~\ref{fig:simple_arpes}{\bf a.}) at small $\sigma_\omega$ the intensity is nearly all from
the (incoherent) populations. As the probe width increases, the signal acquires and oscillatory component (\ref{fig:simple_arpes}{\bf b.}),
which eventually dominates (\ref{fig:simple_arpes}{\bf c.}).
When the photoemission process favors the lower energy state ($\theta=\pi/6$), the majority of the intensity lies in the
lower energy level $\ket{0}$, resulting in an asymmetric spectrum.
A hallmark of {\it coherent} 
state dynamics is that in contrast to other dynamics that give rise to an oscillation in the photoelectron intensity, 
e.g. coherent phonons, there is a variation in intensity without any shifting in energy levels.

\begin{figure}[t]
	\includegraphics[width=0.49\textwidth]{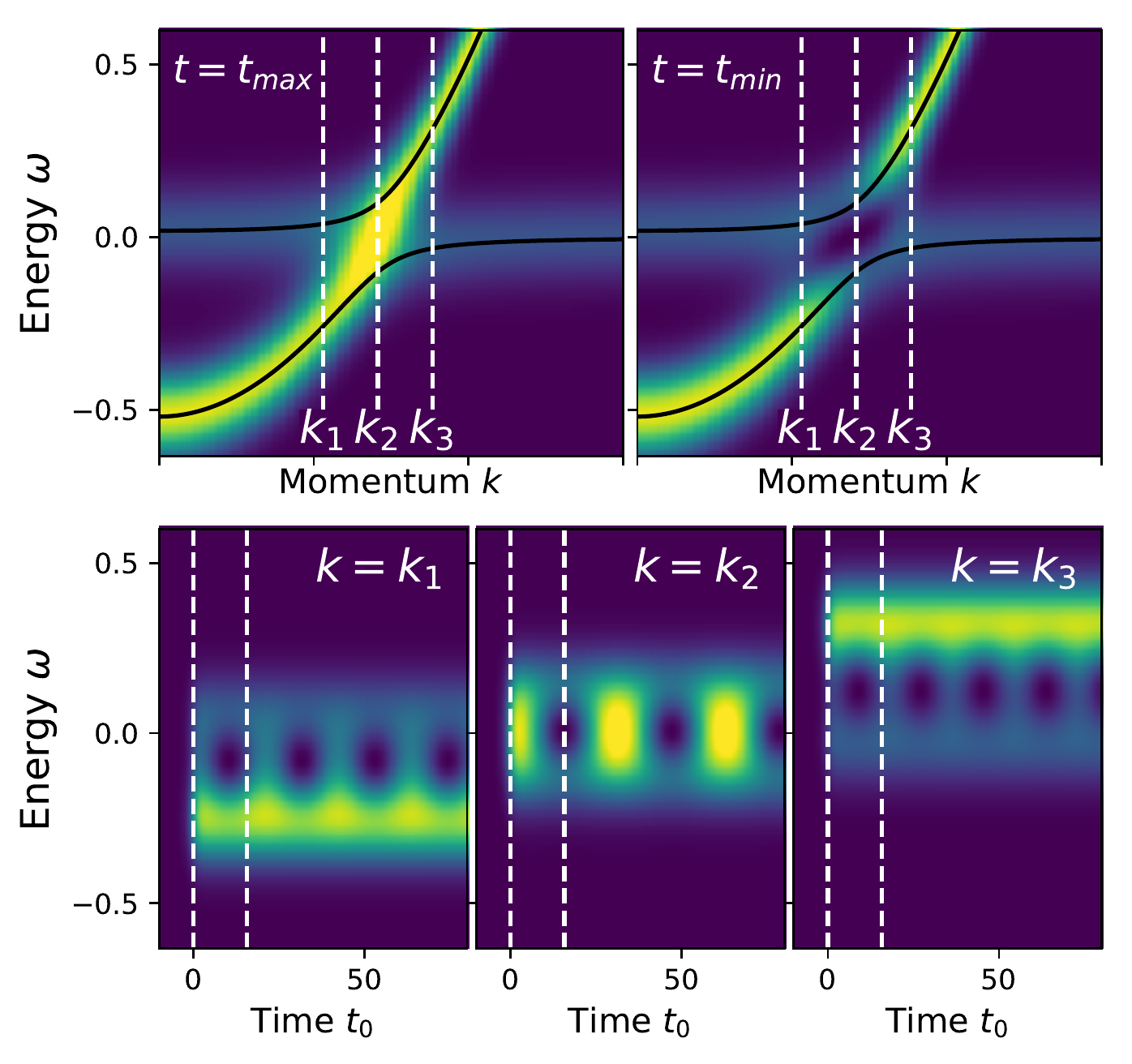}
	\caption{{\bf Time-resolved ARPES of a coherence at an avoided crossing.} 
	The top row shows the tr-ARPES signal at the first maximum and minimum at the gap momentum; the solid lines indicate
	the dispersions.
	The bottom row show energy cuts at the momenta $k_1$ through $k_3$ indicated in the top rows.
	Here, the probe energy resolution is larger than the gap ($1.2\Delta$)
	}
	\label{fig:band_coh}
\end{figure}

\vspace{1em}
\noindent{\bf {Coherences in solids.}}
Moving beyond the simple 2-state system, we can apply the same concepts to coherent states in a band of electrons.  Since the in-plane momentum $\kk$ is a good quantum number
for systems with translation invariance (the dipole transitions happen at $\qq=0$), the
coherence must initially exist between two states with the same value of $\kk$. 
Thus, two states that lie nearby
in energy at the same momentum (and thus are in separate bands) are ideal for this technique. 
Extending the notation above of states $\ket{0}$ and $\ket{1}$ to now indicate electronic bands $a$ and
$b$, the lesser
Green's function for a coherence in momentum space becomes
\begin{align}
G_\kk^{<,\mathrm{coh}}(t,t') &= i \sin(2\theta_\kk) 
\mu_{ca,\kk} \mu_{bc,\kk} e^{-i\left(\frac{\varepsilon_{a,\kk} + \varepsilon_{b,\kk}}{2}\right)t_\mathrm{rel} } \nonumber\\
&\times \cos(\left(\varepsilon_{a,\kk} - \varepsilon_{b,\kk}\right)t_\mathrm{ave}) .
 \end{align}
The momentum-sensitivity of tr-ARPES
also highlights a novel aspect: this measurement allows for the momentum resolution of coherences. We demonstrate the potential of this technique for a model band structure with hybridization gap $\Delta$
between a heavy and a light band. We assume that the bands may be equally populated and photoexcited
at any momentum, but with the photoemission matrix elements preferentially selecting the light band,
and evaluate Eq.~\ref{eq:exact}.
The results are shown in Fig.~\ref{fig:band_coh}, where we plot the dispersions at the minima
and maxima of the oscillatory spectral weight, as well as cuts at fixed momentum as a function of
time, using a probe energy resolution slightly larger than the gap energy ($\sigma_\omega = 1.2\Delta$).
The spectral weight oscillates most strongly where the gap is smallest, i.e. right at the
band crossing in the absence of a hybridization gap.  As we move in momentum away from the maximum, the oscillation
frequency increases as it is equal to the band separation. The matrix elements continue to highlight the contribution
of the light band. These results underscore the potential advantage of
this technique in accessing the smallest energy scales; the energy resolution is one of the inherent limitations
in time-resolved ARPES, yet here it is used to reveal the smallest energy scales by turning to the time domain.
For dispersive bands, the oscillation frequency varies rapidly with momentum, leading to a beating
pattern even though each
momentum oscillates at its own frequency. In turn, this suggests that a
potential inversion is possible; if the oscillations can be measured as a function of momentum, the
gap between the bands may also be resolved as a function of momentum.  Conversely,
if little to no momentum dependence is seen, this suggests a large regime of bands with a constant gap
(e.g. as seen in SmB$_6$\cite{sakhya2020ground}).

\vspace{1em}
\noindent{\bf Decay and Dephasing.}
One major point that remains un-addressed is the decay of populations and coherences.
There are, however, several: population decay rates in $\dyad{0}{0}$-like states,
and decoherence in both the $\dyad{0}{v}$ and $\dyad{0}{1}$-like states.

First, let us consider the population portion of the density matrix.
While the system
is in a population $\dyad{0}{0}$, the decay is solely due to the population transfer out of this state, and this
is typically measured in time-resolved photoemission studies. Naively, in reading the diagram
we would assign a decay factor
$\exp\left(-\Gamma_0 \mathrm{min}\left(t,t'\right)\right)$ since the coherence is present until the earlier of
times $t$ and $t'$. However, this is an oversimplification; the correct approach is to solve the Lindblad
equation (or the time domain Dyson equation\cite{kemper2018general}) because the time dynamics in
average and relative times are not separable.
Here, we will work with an approximation where the population decay rate $\Gamma_0$ is
small compared to the system energy scales ($\Gamma \ll \ea, \Sigma$); this simplifies the decay factor to
$\exp\left(-\Gamma_0 \tave\right)$. 
After the first photoemission operator, the
system is in a $\dyad{v}{0}$ state. This is in principle also a coherence, although it is between an empty
and filled state; this state is the usual one involved in the propagation of a single-particle excitation, and
thus is subject to decay due to the imaginary part of the self-energy ($\Sigma^{''}_0$).  Below,
we will assume that the real part of the self-energy has been absorbed into the quasiparticle energy. 
Thus, the photoemission signal
from the populations $\dyad{0}{0}$ and $\dyad{1}{1}$ becomes
\begin{align}
G^{<,\mathrm{pop}}(t,t') &= i\rho_0^2 e^{-i\ea\trel} e^{-\Gamma_0 \tave} e^{-\Sigma^{''}_0|\trel|}\nonumber\\
&+i\rho_1^2 e^{-i\eb\trel} e^{-\Gamma_1 \tave} e^{-\Sigma^{''}_1|\trel|}
\end{align}
The relationship between $\Gamma_{0/1}$ and $\Sigma^{''}_{0/1}$ is complex, and has been studied in some
detail previously\cite{kemper2018general}; however, both arise due to interactions.

The decay of the coherence has two contributions.
First, the populations in the
levels $\ket{0}$ and $\ket{1}$ decay with their individual rates $\Gamma_0$ and $\Gamma_1$. Second,
there is a ``proper dephasing rate'' that does not involve population decay, which
we denote as $\gamma^\mathrm{prop}_{01}$. The overall rate of decay of the coherence
is given by
\begin{align}
\gamma^\mathrm{coh} = \frac{1}{2}\left(\Gamma_0 + \Gamma_1\right) + \gamma^\mathrm{prop}_{01}.
\label{eq:gamma_coh}
\end{align}
This rate applies while the coherence exists, i.e. during the time interval $0< t,t' < \tave$.
Thus, the coherence decays in a similar fashion as the populations,
but with the decay rate $\Gamma$ given by Eq.~\ref{eq:gamma_coh}.
The influence of the self-energy is more complex because, in principle, the self-energy
can be different for the $\ket{0}$ and $\ket{1}$. If they are identical, $\Sigma_0 =\Sigma_1 = \Sigma$,
then
\begin{align}
G^{<,\mathrm{coh}}(t,t') 
&=  2 i \rho_0 \rho_1 \cos\left(\left(\epsa - \epsb\right) \tave\right) e^{-\gamma^\mathrm{coh} \tave}\nonumber\\
&\times  e^{-i\left(\frac{\epsa + \epsb}{2}\right)\trel} 
e^{-\Sigma^{''} |\trel|}.
\end{align}
The extension to the case where the self-energies are different is straightforward.

\vspace{1em}
\noindent{\bf {\large Discussion}}\\
In this work, we have demonstrated the signatures of optical pump-induced coherences in time-resolved photoemission. Contrary to conventional wisdom where coherences do not contribute to photocurrent, we presented an intriguing case where it does. We considered composite bands composed of different orbitals where the orbital selectivity of photoemission operator is an enabling factor for coherence observation. To evaluate the contributions to photocurrent, we developed a density matrix based two-sided Feynman diagrammatic formalism. We applied the developed formalism to a simple 4-level toy model (qubit with ancillary states) highlighting the factors governing the interference exhibited due to coherence between proximal energy levels in photoemission spectrum. Such manifestations of coherence in photoemission spectrum persists even in presence of decay mechanisms as long as the decay timescale is long compared to the interference beating timescale. Beyond the toy model, we applied the presented formalism to solid-state systems where coherence can be observed at avoided crossings. Thus, our work identified the conditions under which coherences can be observed in time-resolved photoemission measurements. 

The presented work helps lay the foundation for observing signatures of coherence in time-resolved photoemission measurements. It further opens up new avenues for research in a variety of directions. First, we have proposed composite bands composed of different orbitals as a platform for observing such coherences in time-resolved photoemission, where the orbital selectivity of photoemission operator is critical. Evaluating the orbital selectivity from the different orbitals forming the composite bands in various material platforms is an important direction that requires exploration. Second, the next step here is to couple the presented formalism with the Lindblad master equation for open quantum systems to formally and systematically incorporate the decay mechanisms that can hinder coherence observations. And last, identify how such observations of coherences can help advance our understanding of material properties. 

\vspace{1em}
\noindent{\bf {\large Methods.}}\\
\noindent{\bf Time-resolved ARPES.}\\
We make use of the
formalism for time- and angle-resolved photoemission (tr-ARPES) as laid
out by Freericks et al.\cite{freericks_theoretical_2009} They conclude that
tr-ARPES effectively measures an averaged lesser Green's function $G_\kk^<(t,t')$
for each momentum $\kk$;
given a probe that is temporally described by $s(t)$ which is centered
around the measurement time $t_0$, the photocurrent 
$\mathcal{I}(\kk,\omega,t_0)$ may
be written as
\begin{align}
    \mathcal{I}(\kk,\omega,t_0) = -i\iint \dd{t} \dd{t'} s(t) s(t') e^{i\omega (t-t')} G_\kk^<(t,t').
    \label{eq:fourier}
\end{align}
This expression denotes an effective averaging of $G_\kk^<(t,t')$ in a window
set by the probe pulses, and a Fourier transform along the {\it relative}
time direction $\trel \equiv t-t'$. $t_0$ is also known as the {\it average}
time $\tave \equiv \left(1/2\right)\left(t+t'\right)$, and it indicates the time
delay in the experiment between the pump and the probe. Thus, to investigate
tr-ARPES, one has simply to obtain the lesser Green's function. This approach
was used successfully for a variety of systems, including strongly correlated
materials, superconductors, excitonic insulators, as well as simpler interacting systems.\cite{freericks_theoretical_2009,RevModPhys.86.779,*PhysRevB.101.041201,sentef_13,*kemper_effect_2014,*kemper_direct_2015}
These approaches used a Green's function formalism, which naturally provides
access to $G^<(t,t')$. However, coherences are more naturally described by a density
matrix formalism, which is the approach we will follow here.

The lesser Green's function is
\begin{align}
   G_\kk^<(t,t') = i  \langle \hat c_\kk^\dagger(t') \hat c_\kk(t) \rangle.
\end{align}
We may evaluate this expression using a density matrix formalism,
and using  the time evolution
operators $\hat U(t,t')$ for the operators,
\begin{align}
G_\kk^<(t,t') = i \Tr\lbrace \hat U(0,t') \hat c_\kk^\dagger \hat U(t',t) \hat c_\kk \hat U(t,0) \rho \rbrace,
\label{eq:gless_from_rho}
\end{align}
where $\rho$ is the density matrix of the system at a reference time.
In equilibrium
and when single-particle excitations are diagonal, a straightforward evaluation yields
\begin{align}
G_\kk^<(t,t') = i n_\kk e^{-i \varepsilon_\kk (t-t')},
\end{align}
where $n_\kk$ is the Fermi function $n_\kk = \left[ 1 + e^{\beta \varepsilon_\kk}\right]^{-1}$.
To obtain the ARPES spectrum, 
we may simply
rotate $t-t'\rightarrow \trel$ (the relative time coordinate) and perform a Fourier transform from $\trel$ to $\omega$, which would yield a peak at the energy $\omega=\varepsilon_\kk$ and no average time dependence. \\

\bibliography{CoherentARPES_paper}

\begin{thebibliography}{30}%
\makeatletter
\providecommand \@ifxundefined [1]{%
 \@ifx{#1\undefined}
}%
\providecommand \@ifnum [1]{%
 \ifnum #1\expandafter \@firstoftwo
 \else \expandafter \@secondoftwo
 \fi
}%
\providecommand \@ifx [1]{%
 \ifx #1\expandafter \@firstoftwo
 \else \expandafter \@secondoftwo
 \fi
}%
\providecommand \natexlab [1]{#1}%
\providecommand \enquote  [1]{``#1''}%
\providecommand \bibnamefont  [1]{#1}%
\providecommand \bibfnamefont [1]{#1}%
\providecommand \citenamefont [1]{#1}%
\providecommand \href@noop [0]{\@secondoftwo}%
\providecommand \href [0]{\begingroup \@sanitize@url \@href}%
\providecommand \@href[1]{\@@startlink{#1}\@@href}%
\providecommand \@@href[1]{\endgroup#1\@@endlink}%
\providecommand \@sanitize@url [0]{\catcode `\\12\catcode `\$12\catcode
  `\&12\catcode `\#12\catcode `\^12\catcode `\_12\catcode `\%12\relax}%
\providecommand \@@startlink[1]{}%
\providecommand \@@endlink[0]{}%
\providecommand \url  [0]{\begingroup\@sanitize@url \@url }%
\providecommand \@url [1]{\endgroup\@href {#1}{\urlprefix }}%
\providecommand \urlprefix  [0]{URL }%
\providecommand \Eprint [0]{\href }%
\providecommand \doibase [0]{http://dx.doi.org/}%
\providecommand \selectlanguage [0]{\@gobble}%
\providecommand \bibinfo  [0]{\@secondoftwo}%
\providecommand \bibfield  [0]{\@secondoftwo}%
\providecommand \translation [1]{[#1]}%
\providecommand \BibitemOpen [0]{}%
\providecommand \bibitemStop [0]{}%
\providecommand \bibitemNoStop [0]{.\EOS\space}%
\providecommand \EOS [0]{\spacefactor3000\relax}%
\providecommand \BibitemShut  [1]{\csname bibitem#1\endcsname}%
\let\auto@bib@innerbib\@empty
\bibitem [{\citenamefont {{U. S. Department of Energy, Office of Science Basic
  Energy Science Report (2017)}}()}]{DOEQUANTUMPLATFORM}%
  \BibitemOpen
  \bibfield  {author} {\bibinfo {author} {\bibnamefont {{U. S. Department of
  Energy, Office of Science Basic Energy Science Report (2017)}}},\ }\href@noop
  {} {\enquote {\bibinfo {title} {{Opportunities for Basic Research for
  Next-Generation Quantum Systems}},}\ }\bibinfo {note} {Available at:
  \url{https://science.osti.gov/bes/community-resources/reports/}}\BibitemShut
  {NoStop}%
\bibitem [{\citenamefont {Hashimoto}\ \emph {et~al.}(2014)\citenamefont
  {Hashimoto}, \citenamefont {Vishik}, \citenamefont {He}, \citenamefont
  {Devereaux},\ and\ \citenamefont {Shen}}]{hashimoto2014energy}%
  \BibitemOpen
  \bibfield  {author} {\bibinfo {author} {\bibfnamefont {M.}~\bibnamefont
  {Hashimoto}}, \bibinfo {author} {\bibfnamefont {I.~M.}\ \bibnamefont
  {Vishik}}, \bibinfo {author} {\bibfnamefont {R.-H.}\ \bibnamefont {He}},
  \bibinfo {author} {\bibfnamefont {T.~P.}\ \bibnamefont {Devereaux}}, \ and\
  \bibinfo {author} {\bibfnamefont {Z.-X.}\ \bibnamefont {Shen}},\ }\href@noop
  {} {\bibfield  {journal} {\bibinfo  {journal} {Nature Physics}\ }\textbf
  {\bibinfo {volume} {10}},\ \bibinfo {pages} {483} (\bibinfo {year}
  {2014})}\BibitemShut {NoStop}%
\bibitem [{\citenamefont {Demsar}\ \emph {et~al.}(2006)\citenamefont {Demsar},
  \citenamefont {Thorsm\o{}lle}, \citenamefont {Sarrao},\ and\ \citenamefont
  {Taylor}}]{demsar2006photoexcited}%
  \BibitemOpen
  \bibfield  {author} {\bibinfo {author} {\bibfnamefont {J.}~\bibnamefont
  {Demsar}}, \bibinfo {author} {\bibfnamefont {V.~K.}\ \bibnamefont
  {Thorsm\o{}lle}}, \bibinfo {author} {\bibfnamefont {J.~L.}\ \bibnamefont
  {Sarrao}}, \ and\ \bibinfo {author} {\bibfnamefont {A.~J.}\ \bibnamefont
  {Taylor}},\ }\href {\doibase 10.1103/PhysRevLett.96.037401} {\bibfield
  {journal} {\bibinfo  {journal} {Phys. Rev. Lett.}\ }\textbf {\bibinfo
  {volume} {96}},\ \bibinfo {pages} {037401} (\bibinfo {year}
  {2006})}\BibitemShut {NoStop}%
\bibitem [{\citenamefont {Gerber}\ \emph {et~al.}(2017)\citenamefont {Gerber},
  \citenamefont {Yang}, \citenamefont {Zhu}, \citenamefont {Soifer},
  \citenamefont {Sobota}, \citenamefont {Rebec}, \citenamefont {Lee},
  \citenamefont {Jia}, \citenamefont {Moritz}, \citenamefont {Jia} \emph
  {et~al.}}]{gerber2017femtosecond}%
  \BibitemOpen
  \bibfield  {author} {\bibinfo {author} {\bibfnamefont {S.}~\bibnamefont
  {Gerber}}, \bibinfo {author} {\bibfnamefont {S.-L.}\ \bibnamefont {Yang}},
  \bibinfo {author} {\bibfnamefont {D.}~\bibnamefont {Zhu}}, \bibinfo {author}
  {\bibfnamefont {H.}~\bibnamefont {Soifer}}, \bibinfo {author} {\bibfnamefont
  {J.}~\bibnamefont {Sobota}}, \bibinfo {author} {\bibfnamefont
  {S.}~\bibnamefont {Rebec}}, \bibinfo {author} {\bibfnamefont
  {J.}~\bibnamefont {Lee}}, \bibinfo {author} {\bibfnamefont {T.}~\bibnamefont
  {Jia}}, \bibinfo {author} {\bibfnamefont {B.}~\bibnamefont {Moritz}},
  \bibinfo {author} {\bibfnamefont {C.}~\bibnamefont {Jia}},  \emph {et~al.},\
  }\href@noop {} {\bibfield  {journal} {\bibinfo  {journal} {Science}\ }\textbf
  {\bibinfo {volume} {357}},\ \bibinfo {pages} {71} (\bibinfo {year}
  {2017})}\BibitemShut {NoStop}%
\bibitem [{\citenamefont {Hein}\ \emph {et~al.}(2019)\citenamefont {Hein},
  \citenamefont {Jauernik}, \citenamefont {Erk}, \citenamefont {Yang},
  \citenamefont {Qi}, \citenamefont {Sun}, \citenamefont {Felser},\ and\
  \citenamefont {Bauer}}]{hein2019mode}%
  \BibitemOpen
  \bibfield  {author} {\bibinfo {author} {\bibfnamefont {P.}~\bibnamefont
  {Hein}}, \bibinfo {author} {\bibfnamefont {S.}~\bibnamefont {Jauernik}},
  \bibinfo {author} {\bibfnamefont {H.}~\bibnamefont {Erk}}, \bibinfo {author}
  {\bibfnamefont {L.}~\bibnamefont {Yang}}, \bibinfo {author} {\bibfnamefont
  {Y.}~\bibnamefont {Qi}}, \bibinfo {author} {\bibfnamefont {Y.}~\bibnamefont
  {Sun}}, \bibinfo {author} {\bibfnamefont {C.}~\bibnamefont {Felser}}, \ and\
  \bibinfo {author} {\bibfnamefont {M.}~\bibnamefont {Bauer}},\ }\href@noop {}
  {\bibfield  {journal} {\bibinfo  {journal} {arXiv preprint arXiv:1911.12166}\
  } (\bibinfo {year} {2019})}\BibitemShut {NoStop}%
\bibitem [{\citenamefont {Yang}\ \emph {et~al.}(2019)\citenamefont {Yang},
  \citenamefont {Sobota}, \citenamefont {He}, \citenamefont {Leuenberger},
  \citenamefont {Soifer}, \citenamefont {Eisaki}, \citenamefont {Kirchmann},\
  and\ \citenamefont {Shen}}]{yang2019mode}%
  \BibitemOpen
  \bibfield  {author} {\bibinfo {author} {\bibfnamefont {S.-L.}\ \bibnamefont
  {Yang}}, \bibinfo {author} {\bibfnamefont {J.}~\bibnamefont {Sobota}},
  \bibinfo {author} {\bibfnamefont {Y.}~\bibnamefont {He}}, \bibinfo {author}
  {\bibfnamefont {D.}~\bibnamefont {Leuenberger}}, \bibinfo {author}
  {\bibfnamefont {H.}~\bibnamefont {Soifer}}, \bibinfo {author} {\bibfnamefont
  {H.}~\bibnamefont {Eisaki}}, \bibinfo {author} {\bibfnamefont
  {P.}~\bibnamefont {Kirchmann}}, \ and\ \bibinfo {author} {\bibfnamefont
  {Z.-X.}\ \bibnamefont {Shen}},\ }\href@noop {} {\bibfield  {journal}
  {\bibinfo  {journal} {Physical review letters}\ }\textbf {\bibinfo {volume}
  {122}},\ \bibinfo {pages} {176403} (\bibinfo {year} {2019})}\BibitemShut
  {NoStop}%
\bibitem [{\citenamefont {Zhang}\ \emph {et~al.}(2020)\citenamefont {Zhang},
  \citenamefont {Shi}, \citenamefont {You}, \citenamefont {Tao}, \citenamefont
  {Zhong}, \citenamefont {Kabeer}, \citenamefont {Maldonado}, \citenamefont
  {Oppeneer}, \citenamefont {Bauer}, \citenamefont {Rossnagel} \emph
  {et~al.}}]{zhang2020coherent}%
  \BibitemOpen
  \bibfield  {author} {\bibinfo {author} {\bibfnamefont {Y.}~\bibnamefont
  {Zhang}}, \bibinfo {author} {\bibfnamefont {X.}~\bibnamefont {Shi}}, \bibinfo
  {author} {\bibfnamefont {W.}~\bibnamefont {You}}, \bibinfo {author}
  {\bibfnamefont {Z.}~\bibnamefont {Tao}}, \bibinfo {author} {\bibfnamefont
  {Y.}~\bibnamefont {Zhong}}, \bibinfo {author} {\bibfnamefont {F.~C.}\
  \bibnamefont {Kabeer}}, \bibinfo {author} {\bibfnamefont {P.}~\bibnamefont
  {Maldonado}}, \bibinfo {author} {\bibfnamefont {P.~M.}\ \bibnamefont
  {Oppeneer}}, \bibinfo {author} {\bibfnamefont {M.}~\bibnamefont {Bauer}},
  \bibinfo {author} {\bibfnamefont {K.}~\bibnamefont {Rossnagel}},  \emph
  {et~al.},\ }\href@noop {} {\bibfield  {journal} {\bibinfo  {journal}
  {Proceedings of the National Academy of Sciences}\ }\textbf {\bibinfo
  {volume} {117}},\ \bibinfo {pages} {8788} (\bibinfo {year}
  {2020})}\BibitemShut {NoStop}%
\bibitem [{\citenamefont {Kemper}\ \emph {et~al.}(2018)\citenamefont {Kemper},
  \citenamefont {Abdurazakov},\ and\ \citenamefont
  {Freericks}}]{kemper2018general}%
  \BibitemOpen
  \bibfield  {author} {\bibinfo {author} {\bibfnamefont {A.}~\bibnamefont
  {Kemper}}, \bibinfo {author} {\bibfnamefont {O.}~\bibnamefont {Abdurazakov}},
  \ and\ \bibinfo {author} {\bibfnamefont {J.}~\bibnamefont {Freericks}},\
  }\href@noop {} {\bibfield  {journal} {\bibinfo  {journal} {Physical Review
  X}\ }\textbf {\bibinfo {volume} {8}},\ \bibinfo {pages} {041009} (\bibinfo
  {year} {2018})}\BibitemShut {NoStop}%
\bibitem [{\citenamefont {Aoki}\ \emph {et~al.}(2014)\citenamefont {Aoki},
  \citenamefont {Tsuji}, \citenamefont {Eckstein}, \citenamefont {Kollar},
  \citenamefont {Oka},\ and\ \citenamefont {Werner}}]{RevModPhys.86.779}%
  \BibitemOpen
  \bibfield  {author} {\bibinfo {author} {\bibfnamefont {H.}~\bibnamefont
  {Aoki}}, \bibinfo {author} {\bibfnamefont {N.}~\bibnamefont {Tsuji}},
  \bibinfo {author} {\bibfnamefont {M.}~\bibnamefont {Eckstein}}, \bibinfo
  {author} {\bibfnamefont {M.}~\bibnamefont {Kollar}}, \bibinfo {author}
  {\bibfnamefont {T.}~\bibnamefont {Oka}}, \ and\ \bibinfo {author}
  {\bibfnamefont {P.}~\bibnamefont {Werner}},\ }\href {\doibase
  10.1103/RevModPhys.86.779} {\bibfield  {journal} {\bibinfo  {journal} {Rev.
  Mod. Phys.}\ }\textbf {\bibinfo {volume} {86}},\ \bibinfo {pages} {779}
  (\bibinfo {year} {2014})}\BibitemShut {NoStop}%
\bibitem [{\citenamefont {{Rameau}}\ \emph {et~al.}(2016)\citenamefont
  {{Rameau}}, \citenamefont {{Freutel}}, \citenamefont {{Kemper}},
  \citenamefont {{Sentef}}, \citenamefont {{Freericks}}, \citenamefont
  {{Avigo}}, \citenamefont {{Ligges}}, \citenamefont {{Rettig}}, \citenamefont
  {{Yoshida}}, \citenamefont {{Eisaki}}, \citenamefont {{Schneeloch}},
  \citenamefont {{Zhong}}, \citenamefont {{Xu}}, \citenamefont {{Gu}},
  \citenamefont {{Johnson}},\ and\ \citenamefont
  {{Bovensiepen}}}]{rameau_energy_2016}%
  \BibitemOpen
  \bibfield  {author} {\bibinfo {author} {\bibfnamefont {J.~D.}\ \bibnamefont
  {{Rameau}}}, \bibinfo {author} {\bibfnamefont {S.}~\bibnamefont {{Freutel}}},
  \bibinfo {author} {\bibfnamefont {A.~F.}\ \bibnamefont {{Kemper}}}, \bibinfo
  {author} {\bibfnamefont {M.~A.}\ \bibnamefont {{Sentef}}}, \bibinfo {author}
  {\bibfnamefont {J.~K.}\ \bibnamefont {{Freericks}}}, \bibinfo {author}
  {\bibfnamefont {I.}~\bibnamefont {{Avigo}}}, \bibinfo {author} {\bibfnamefont
  {M.}~\bibnamefont {{Ligges}}}, \bibinfo {author} {\bibfnamefont
  {L.}~\bibnamefont {{Rettig}}}, \bibinfo {author} {\bibfnamefont
  {Y.}~\bibnamefont {{Yoshida}}}, \bibinfo {author} {\bibfnamefont
  {H.}~\bibnamefont {{Eisaki}}}, \bibinfo {author} {\bibfnamefont
  {J.}~\bibnamefont {{Schneeloch}}}, \bibinfo {author} {\bibfnamefont {R.~D.}\
  \bibnamefont {{Zhong}}}, \bibinfo {author} {\bibfnamefont {Z.~J.}\
  \bibnamefont {{Xu}}}, \bibinfo {author} {\bibfnamefont {G.~D.}\ \bibnamefont
  {{Gu}}}, \bibinfo {author} {\bibfnamefont {P.~D.}\ \bibnamefont {{Johnson}}},
  \ and\ \bibinfo {author} {\bibfnamefont {U.}~\bibnamefont {{Bovensiepen}}},\
  }\href {\doibase 10.1038/ncomms13761} {\bibfield  {journal} {\bibinfo
  {journal} {Nature Communications}\ }\textbf {\bibinfo {volume} {7}},\
  \bibinfo {eid} {13761} (\bibinfo {year} {2016})}\BibitemShut {NoStop}%
\bibitem [{\citenamefont {Konstantinova}\ \emph {et~al.}(2018)\citenamefont
  {Konstantinova}, \citenamefont {Rameau}, \citenamefont {Reid}, \citenamefont
  {Abdurazakov}, \citenamefont {Wu}, \citenamefont {Li}, \citenamefont {Shen},
  \citenamefont {Gu}, \citenamefont {Huang}, \citenamefont {Rettig} \emph
  {et~al.}}]{konstantinova2018nonequilibrium}%
  \BibitemOpen
  \bibfield  {author} {\bibinfo {author} {\bibfnamefont {T.}~\bibnamefont
  {Konstantinova}}, \bibinfo {author} {\bibfnamefont {J.~D.}\ \bibnamefont
  {Rameau}}, \bibinfo {author} {\bibfnamefont {A.~H.}\ \bibnamefont {Reid}},
  \bibinfo {author} {\bibfnamefont {O.}~\bibnamefont {Abdurazakov}}, \bibinfo
  {author} {\bibfnamefont {L.}~\bibnamefont {Wu}}, \bibinfo {author}
  {\bibfnamefont {R.}~\bibnamefont {Li}}, \bibinfo {author} {\bibfnamefont
  {X.}~\bibnamefont {Shen}}, \bibinfo {author} {\bibfnamefont {G.}~\bibnamefont
  {Gu}}, \bibinfo {author} {\bibfnamefont {Y.}~\bibnamefont {Huang}}, \bibinfo
  {author} {\bibfnamefont {L.}~\bibnamefont {Rettig}},  \emph {et~al.},\
  }\href@noop {} {\bibfield  {journal} {\bibinfo  {journal} {Science advances}\
  }\textbf {\bibinfo {volume} {4}},\ \bibinfo {pages} {eaap7427} (\bibinfo
  {year} {2018})}\BibitemShut {NoStop}%
\bibitem [{\citenamefont {Freutel}\ \emph {et~al.}(2019)\citenamefont
  {Freutel}, \citenamefont {Rameau}, \citenamefont {Rettig}, \citenamefont
  {Avigo}, \citenamefont {Ligges}, \citenamefont {Yoshida}, \citenamefont
  {Eisaki}, \citenamefont {Schneeloch}, \citenamefont {Zhong}, \citenamefont
  {Xu} \emph {et~al.}}]{freutel2019optical}%
  \BibitemOpen
  \bibfield  {author} {\bibinfo {author} {\bibfnamefont {S.}~\bibnamefont
  {Freutel}}, \bibinfo {author} {\bibfnamefont {J.}~\bibnamefont {Rameau}},
  \bibinfo {author} {\bibfnamefont {L.}~\bibnamefont {Rettig}}, \bibinfo
  {author} {\bibfnamefont {I.}~\bibnamefont {Avigo}}, \bibinfo {author}
  {\bibfnamefont {M.}~\bibnamefont {Ligges}}, \bibinfo {author} {\bibfnamefont
  {Y.}~\bibnamefont {Yoshida}}, \bibinfo {author} {\bibfnamefont
  {H.}~\bibnamefont {Eisaki}}, \bibinfo {author} {\bibfnamefont
  {J.}~\bibnamefont {Schneeloch}}, \bibinfo {author} {\bibfnamefont
  {R.}~\bibnamefont {Zhong}}, \bibinfo {author} {\bibfnamefont
  {Z.}~\bibnamefont {Xu}},  \emph {et~al.},\ }\href@noop {} {\bibfield
  {journal} {\bibinfo  {journal} {Physical Review B}\ }\textbf {\bibinfo
  {volume} {99}},\ \bibinfo {pages} {081116} (\bibinfo {year}
  {2019})}\BibitemShut {NoStop}%
\bibitem [{\citenamefont {Boyd}(2003)}]{boyd2003nonlinear}%
  \BibitemOpen
  \bibfield  {author} {\bibinfo {author} {\bibfnamefont {R.~W.}\ \bibnamefont
  {Boyd}},\ }\href@noop {} {\emph {\bibinfo {title} {Nonlinear optics}}}\
  (\bibinfo  {publisher} {Elsevier},\ \bibinfo {year} {2003})\BibitemShut
  {NoStop}%
\bibitem [{\citenamefont {Moody}\ and\ \citenamefont
  {Cundiff}(2017)}]{moody2017advances}%
  \BibitemOpen
  \bibfield  {author} {\bibinfo {author} {\bibfnamefont {G.}~\bibnamefont
  {Moody}}\ and\ \bibinfo {author} {\bibfnamefont {S.~T.}\ \bibnamefont
  {Cundiff}},\ }\href@noop {} {\bibfield  {journal} {\bibinfo  {journal}
  {Advances in physics: X}\ }\textbf {\bibinfo {volume} {2}},\ \bibinfo {pages}
  {641} (\bibinfo {year} {2017})}\BibitemShut {NoStop}%
\bibitem [{\citenamefont {Kandada}\ and\ \citenamefont
  {Silva}(2019)}]{kandada2019perspective}%
  \BibitemOpen
  \bibfield  {author} {\bibinfo {author} {\bibfnamefont {A.~R.~S.}\
  \bibnamefont {Kandada}}\ and\ \bibinfo {author} {\bibfnamefont
  {C.}~\bibnamefont {Silva}},\ }\href@noop {} {\bibfield  {journal} {\bibinfo
  {journal} {arXiv preprint arXiv:1908.03909}\ } (\bibinfo {year}
  {2019})}\BibitemShut {NoStop}%
\bibitem [{\citenamefont {Thouin}\ \emph {et~al.}(2019)\citenamefont {Thouin},
  \citenamefont {Cortecchia}, \citenamefont {Petrozza}, \citenamefont
  {Kandada},\ and\ \citenamefont {Silva}}]{thouin2019enhanced}%
  \BibitemOpen
  \bibfield  {author} {\bibinfo {author} {\bibfnamefont {F.}~\bibnamefont
  {Thouin}}, \bibinfo {author} {\bibfnamefont {D.}~\bibnamefont {Cortecchia}},
  \bibinfo {author} {\bibfnamefont {A.}~\bibnamefont {Petrozza}}, \bibinfo
  {author} {\bibfnamefont {A.~R.~S.}\ \bibnamefont {Kandada}}, \ and\ \bibinfo
  {author} {\bibfnamefont {C.}~\bibnamefont {Silva}},\ }\href@noop {}
  {\bibfield  {journal} {\bibinfo  {journal} {Physical Review Research}\
  }\textbf {\bibinfo {volume} {1}},\ \bibinfo {pages} {032032} (\bibinfo {year}
  {2019})}\BibitemShut {NoStop}%
\bibitem [{\citenamefont {Smallwood}\ and\ \citenamefont
  {Cundiff}(2018)}]{smallwood2018multidimensional}%
  \BibitemOpen
  \bibfield  {author} {\bibinfo {author} {\bibfnamefont {C.~L.}\ \bibnamefont
  {Smallwood}}\ and\ \bibinfo {author} {\bibfnamefont {S.~T.}\ \bibnamefont
  {Cundiff}},\ }\href@noop {} {\bibfield  {journal} {\bibinfo  {journal} {Laser
  \& Photonics Reviews}\ }\textbf {\bibinfo {volume} {12}},\ \bibinfo {pages}
  {1800171} (\bibinfo {year} {2018})}\BibitemShut {NoStop}%
\bibitem [{\citenamefont {Yee}\ \emph {et~al.}(1977)\citenamefont {Yee},
  \citenamefont {Gustafson}, \citenamefont {Druet},\ and\ \citenamefont
  {Taran}}]{yee1977diagrammatic}%
  \BibitemOpen
  \bibfield  {author} {\bibinfo {author} {\bibfnamefont {S.}~\bibnamefont
  {Yee}}, \bibinfo {author} {\bibfnamefont {T.}~\bibnamefont {Gustafson}},
  \bibinfo {author} {\bibfnamefont {S.}~\bibnamefont {Druet}}, \ and\ \bibinfo
  {author} {\bibfnamefont {J.-P.}\ \bibnamefont {Taran}},\ }\href@noop {}
  {\bibfield  {journal} {\bibinfo  {journal} {Optics Communications}\ }\textbf
  {\bibinfo {volume} {23}},\ \bibinfo {pages} {1} (\bibinfo {year}
  {1977})}\BibitemShut {NoStop}%
\bibitem [{\citenamefont {Yee}\ and\ \citenamefont
  {Gustafson}(1978)}]{yee1978diagrammatic}%
  \BibitemOpen
  \bibfield  {author} {\bibinfo {author} {\bibfnamefont {T.}~\bibnamefont
  {Yee}}\ and\ \bibinfo {author} {\bibfnamefont {T.}~\bibnamefont
  {Gustafson}},\ }\href@noop {} {\bibfield  {journal} {\bibinfo  {journal}
  {Physical Review A}\ }\textbf {\bibinfo {volume} {18}},\ \bibinfo {pages}
  {1597} (\bibinfo {year} {1978})}\BibitemShut {NoStop}%
\bibitem [{\citenamefont {Gordon}\ \emph {et~al.}(2013)\citenamefont {Gordon},
  \citenamefont {Weber}, \citenamefont {Varley}, \citenamefont {Janotti},
  \citenamefont {Awschalom},\ and\ \citenamefont {Van~de
  Walle}}]{gordon2013quantum}%
  \BibitemOpen
  \bibfield  {author} {\bibinfo {author} {\bibfnamefont {L.}~\bibnamefont
  {Gordon}}, \bibinfo {author} {\bibfnamefont {J.~R.}\ \bibnamefont {Weber}},
  \bibinfo {author} {\bibfnamefont {J.~B.}\ \bibnamefont {Varley}}, \bibinfo
  {author} {\bibfnamefont {A.}~\bibnamefont {Janotti}}, \bibinfo {author}
  {\bibfnamefont {D.~D.}\ \bibnamefont {Awschalom}}, \ and\ \bibinfo {author}
  {\bibfnamefont {C.~G.}\ \bibnamefont {Van~de Walle}},\ }\href@noop {}
  {\bibfield  {journal} {\bibinfo  {journal} {MRS bulletin}\ }\textbf {\bibinfo
  {volume} {38}},\ \bibinfo {pages} {802} (\bibinfo {year} {2013})}\BibitemShut
  {NoStop}%
\bibitem [{\citenamefont {Hays}\ \emph {et~al.}(2019)\citenamefont {Hays},
  \citenamefont {Fatemi}, \citenamefont {Serniak}, \citenamefont {Bouman},
  \citenamefont {Diamond}, \citenamefont {de~Lange}, \citenamefont {Krogstrup},
  \citenamefont {Nyg{\aa}rd}, \citenamefont {Geresdi},\ and\ \citenamefont
  {Devoret}}]{hays2019continuous}%
  \BibitemOpen
  \bibfield  {author} {\bibinfo {author} {\bibfnamefont {M.}~\bibnamefont
  {Hays}}, \bibinfo {author} {\bibfnamefont {V.}~\bibnamefont {Fatemi}},
  \bibinfo {author} {\bibfnamefont {K.}~\bibnamefont {Serniak}}, \bibinfo
  {author} {\bibfnamefont {D.}~\bibnamefont {Bouman}}, \bibinfo {author}
  {\bibfnamefont {S.}~\bibnamefont {Diamond}}, \bibinfo {author} {\bibfnamefont
  {G.}~\bibnamefont {de~Lange}}, \bibinfo {author} {\bibfnamefont
  {P.}~\bibnamefont {Krogstrup}}, \bibinfo {author} {\bibfnamefont
  {J.}~\bibnamefont {Nyg{\aa}rd}}, \bibinfo {author} {\bibfnamefont
  {A.}~\bibnamefont {Geresdi}}, \ and\ \bibinfo {author} {\bibfnamefont
  {M.}~\bibnamefont {Devoret}},\ }\href@noop {} {\bibfield  {journal} {\bibinfo
   {journal} {arXiv:1908.02800}\ } (\bibinfo {year} {2019})}\BibitemShut
  {NoStop}%
\bibitem [{\citenamefont {Gottscholl}\ \emph {et~al.}(2020)\citenamefont
  {Gottscholl}, \citenamefont {Kianinia}, \citenamefont {Soltamov},
  \citenamefont {Orlinskii}, \citenamefont {Mamin}, \citenamefont {Bradac},
  \citenamefont {Kasper}, \citenamefont {Krambrock}, \citenamefont {Sperlich},
  \citenamefont {Toth} \emph {et~al.}}]{gottscholl2020initialization}%
  \BibitemOpen
  \bibfield  {author} {\bibinfo {author} {\bibfnamefont {A.}~\bibnamefont
  {Gottscholl}}, \bibinfo {author} {\bibfnamefont {M.}~\bibnamefont
  {Kianinia}}, \bibinfo {author} {\bibfnamefont {V.}~\bibnamefont {Soltamov}},
  \bibinfo {author} {\bibfnamefont {S.}~\bibnamefont {Orlinskii}}, \bibinfo
  {author} {\bibfnamefont {G.}~\bibnamefont {Mamin}}, \bibinfo {author}
  {\bibfnamefont {C.}~\bibnamefont {Bradac}}, \bibinfo {author} {\bibfnamefont
  {C.}~\bibnamefont {Kasper}}, \bibinfo {author} {\bibfnamefont
  {K.}~\bibnamefont {Krambrock}}, \bibinfo {author} {\bibfnamefont
  {A.}~\bibnamefont {Sperlich}}, \bibinfo {author} {\bibfnamefont
  {M.}~\bibnamefont {Toth}},  \emph {et~al.},\ }\href@noop {} {\bibfield
  {journal} {\bibinfo  {journal} {Nature Materials}\ ,\ \bibinfo {pages} {1}}
  (\bibinfo {year} {2020})}\BibitemShut {NoStop}%
\bibitem [{\citenamefont {Borjans}\ \emph {et~al.}(2020)\citenamefont
  {Borjans}, \citenamefont {Croot}, \citenamefont {Mi}, \citenamefont
  {Gullans},\ and\ \citenamefont {Petta}}]{borjans2020resonant}%
  \BibitemOpen
  \bibfield  {author} {\bibinfo {author} {\bibfnamefont {F.}~\bibnamefont
  {Borjans}}, \bibinfo {author} {\bibfnamefont {X.}~\bibnamefont {Croot}},
  \bibinfo {author} {\bibfnamefont {X.}~\bibnamefont {Mi}}, \bibinfo {author}
  {\bibfnamefont {M.}~\bibnamefont {Gullans}}, \ and\ \bibinfo {author}
  {\bibfnamefont {J.}~\bibnamefont {Petta}},\ }\href@noop {} {\bibfield
  {journal} {\bibinfo  {journal} {Nature}\ }\textbf {\bibinfo {volume} {577}},\
  \bibinfo {pages} {195} (\bibinfo {year} {2020})}\BibitemShut {NoStop}%
\bibitem [{\citenamefont {Lane}\ \emph {et~al.}(2020)\citenamefont {Lane},
  \citenamefont {Tan}, \citenamefont {Beysengulov}, \citenamefont {Nasyedkin},
  \citenamefont {Brook}, \citenamefont {Zhang}, \citenamefont {Stefanski},
  \citenamefont {Byeon}, \citenamefont {Murch},\ and\ \citenamefont
  {Pollanen}}]{lane2020integrating}%
  \BibitemOpen
  \bibfield  {author} {\bibinfo {author} {\bibfnamefont {J.}~\bibnamefont
  {Lane}}, \bibinfo {author} {\bibfnamefont {D.}~\bibnamefont {Tan}}, \bibinfo
  {author} {\bibfnamefont {N.}~\bibnamefont {Beysengulov}}, \bibinfo {author}
  {\bibfnamefont {K.}~\bibnamefont {Nasyedkin}}, \bibinfo {author}
  {\bibfnamefont {E.}~\bibnamefont {Brook}}, \bibinfo {author} {\bibfnamefont
  {L.}~\bibnamefont {Zhang}}, \bibinfo {author} {\bibfnamefont
  {T.}~\bibnamefont {Stefanski}}, \bibinfo {author} {\bibfnamefont
  {H.}~\bibnamefont {Byeon}}, \bibinfo {author} {\bibfnamefont
  {K.}~\bibnamefont {Murch}}, \ and\ \bibinfo {author} {\bibfnamefont
  {J.}~\bibnamefont {Pollanen}},\ }\href@noop {} {\bibfield  {journal}
  {\bibinfo  {journal} {Physical Review A}\ }\textbf {\bibinfo {volume}
  {101}},\ \bibinfo {pages} {012336} (\bibinfo {year} {2020})}\BibitemShut
  {NoStop}%
\bibitem [{\citenamefont {Sakhya}\ and\ \citenamefont
  {Maiti}(2020)}]{sakhya2020ground}%
  \BibitemOpen
  \bibfield  {author} {\bibinfo {author} {\bibfnamefont {A.~P.}\ \bibnamefont
  {Sakhya}}\ and\ \bibinfo {author} {\bibfnamefont {K.}~\bibnamefont {Maiti}},\
  }\href@noop {} {\bibfield  {journal} {\bibinfo  {journal} {Scientific
  Reports}\ }\textbf {\bibinfo {volume} {10}},\ \bibinfo {pages} {1} (\bibinfo
  {year} {2020})}\BibitemShut {NoStop}%
\bibitem [{\citenamefont {Freericks}\ \emph {et~al.}(2009)\citenamefont
  {Freericks}, \citenamefont {Krishnamurthy},\ and\ \citenamefont
  {Pruschke}}]{freericks_theoretical_2009}%
  \BibitemOpen
  \bibfield  {author} {\bibinfo {author} {\bibfnamefont {J.~K.}\ \bibnamefont
  {Freericks}}, \bibinfo {author} {\bibfnamefont {H.~R.}\ \bibnamefont
  {Krishnamurthy}}, \ and\ \bibinfo {author} {\bibfnamefont {T.}~\bibnamefont
  {Pruschke}},\ }\href {\doibase 10.1103/PhysRevLett.102.136401} {\bibfield
  {journal} {\bibinfo  {journal} {Physical Review Letters}\ }\textbf {\bibinfo
  {volume} {102}},\ \bibinfo {pages} {136401} (\bibinfo {year}
  {2009})}\BibitemShut {NoStop}%
\bibitem [{\citenamefont {Perfetto}\ \emph {et~al.}(2020)\citenamefont
  {Perfetto}, \citenamefont {Bianchi},\ and\ \citenamefont
  {Stefanucci}}]{PhysRevB.101.041201}%
  \BibitemOpen
  \bibfield  {author} {\bibinfo {author} {\bibfnamefont {E.}~\bibnamefont
  {Perfetto}}, \bibinfo {author} {\bibfnamefont {S.}~\bibnamefont {Bianchi}}, \
  and\ \bibinfo {author} {\bibfnamefont {G.}~\bibnamefont {Stefanucci}},\
  }\href {\doibase 10.1103/PhysRevB.101.041201} {\bibfield  {journal} {\bibinfo
   {journal} {Phys. Rev. B}\ }\textbf {\bibinfo {volume} {101}},\ \bibinfo
  {pages} {041201} (\bibinfo {year} {2020})}\BibitemShut {NoStop}%
\bibitem [{\citenamefont {Sentef}\ \emph {et~al.}(2013)\citenamefont {Sentef},
  \citenamefont {Kemper}, \citenamefont {Moritz}, \citenamefont {Freericks},
  \citenamefont {Shen},\ and\ \citenamefont {Devereaux}}]{sentef_13}%
  \BibitemOpen
  \bibfield  {author} {\bibinfo {author} {\bibfnamefont {M.~A.}\ \bibnamefont
  {Sentef}}, \bibinfo {author} {\bibfnamefont {A.~F.}\ \bibnamefont {Kemper}},
  \bibinfo {author} {\bibfnamefont {B.}~\bibnamefont {Moritz}}, \bibinfo
  {author} {\bibfnamefont {J.~K.}\ \bibnamefont {Freericks}}, \bibinfo {author}
  {\bibfnamefont {Z.-X.}\ \bibnamefont {Shen}}, \ and\ \bibinfo {author}
  {\bibfnamefont {T.~P.}\ \bibnamefont {Devereaux}},\ }\href@noop {} {\bibfield
   {journal} {\bibinfo  {journal} {Phys. Rev. X}\ }\textbf {\bibinfo {volume}
  {3}},\ \bibinfo {pages} {041033} (\bibinfo {year} {2013})}\BibitemShut
  {NoStop}%
\bibitem [{\citenamefont {Kemper}\ \emph {et~al.}(2014)\citenamefont {Kemper},
  \citenamefont {Sentef}, \citenamefont {Moritz}, \citenamefont {Freericks},\
  and\ \citenamefont {Devereaux}}]{kemper_effect_2014}%
  \BibitemOpen
  \bibfield  {author} {\bibinfo {author} {\bibfnamefont {A.~F.}\ \bibnamefont
  {Kemper}}, \bibinfo {author} {\bibfnamefont {M.~A.}\ \bibnamefont {Sentef}},
  \bibinfo {author} {\bibfnamefont {B.}~\bibnamefont {Moritz}}, \bibinfo
  {author} {\bibfnamefont {J.~K.}\ \bibnamefont {Freericks}}, \ and\ \bibinfo
  {author} {\bibfnamefont {T.~P.}\ \bibnamefont {Devereaux}},\ }\href {\doibase
  10.1103/PhysRevB.90.075126} {\bibfield  {journal} {\bibinfo  {journal} {Phys.
  Rev. B}\ }\textbf {\bibinfo {volume} {90}},\ \bibinfo {pages} {075126}
  (\bibinfo {year} {2014})}\BibitemShut {NoStop}%
\bibitem [{\citenamefont {{Kemper}}\ \emph {et~al.}(2015)\citenamefont
  {{Kemper}}, \citenamefont {{Sentef}}, \citenamefont {{Moritz}}, \citenamefont
  {{Freericks}},\ and\ \citenamefont {{Devereaux}}}]{kemper_direct_2015}%
  \BibitemOpen
  \bibfield  {author} {\bibinfo {author} {\bibfnamefont {A.~F.}\ \bibnamefont
  {{Kemper}}}, \bibinfo {author} {\bibfnamefont {M.~A.}\ \bibnamefont
  {{Sentef}}}, \bibinfo {author} {\bibfnamefont {B.}~\bibnamefont {{Moritz}}},
  \bibinfo {author} {\bibfnamefont {J.~K.}\ \bibnamefont {{Freericks}}}, \ and\
  \bibinfo {author} {\bibfnamefont {T.~P.}\ \bibnamefont {{Devereaux}}},\
  }\href {\doibase 10.1103/PhysRevB.92.224517} {\bibfield  {journal} {\bibinfo
  {journal} {Phys. Rev. B}\ }\textbf {\bibinfo {volume} {92}},\ \bibinfo {eid}
  {224517} (\bibinfo {year} {2015})}\BibitemShut {NoStop}%
\end{thebibliography}%

\begin{acknowledgments}
\noindent We would like to thank P. Kirchmann for helpful discussions.
A.F.K. was supported by NSF DMR-1752713.
\end{acknowledgments}

\section*{Author contributions}
\noindent  A.F.K and A.R. contributed equally to the development of the ideas and the writing of the manuscript. A.F.K. carried
out the mathematical derivations and the development of the diagrammatic language.

\end{document}